\documentstyle[12pt]{article}

\newcommand{\pagenumber}{\pagestyle{plain}\setcounter{page}{1}}

\newcommand{\vc}{\vec{c}}
\newcommand{\z}{\vec{z}}
\newcommand{\zi}{\vec{z}_{i}}
\newcommand{\zj}{\vec{z}_{j}}
\newcommand{\vz}[1]{\vec{z}_{#1}}
\newcommand{\pz}{\vec{z}\,'}
\newcommand{\pzi}{\vec{z}\,'_{i}}
\newcommand{\pzj}{\vec{z}\,'_{j}}
\newcommand{\pvz}[1]{\vec{z}\,'_{#1}}
\newcommand{\norm}[2]{|#1-#2|^{2k_{i}k_{j}}}
\newcommand{\pnorm}[3]{\prod_{#1} |#2-#3|^{2k_{i}k_{j}}}
\newcommand{\msr}[1]{d^{n}z_{#1}\,}
\newcommand{\pmsr}[1]{d^{n}z'_{#1}\,}
\newcommand{\Rt}{R^{T}}

\def\fnote#1#2{\begingroup\def\thefootnote{#1}\footnote{#2}
    \addtocounter{footnote}{-1}\endgroup}
\def\sppt{makoto@sbitp.ucsb.edu}

\begin{document}

\pagestyle{empty}

\vspace{24pt}
\begin{center}
{\bf Natural Generalization of Bosonic String Amplitudes}

\vspace{36pt}
Makoto Natsuume\fnote{*}{\sppt}

\vspace{6pt}
Institute for Theoretical Physics \\
University of California \\
Santa Barbara, California 93106-4030 \\
\vspace{6pt}
and \\
\vspace{6pt}
Theory Group \\ Department of Physics\\
University of Texas \\ Austin, Texas 78712
\vspace{48pt}

\underline{ABSTRACT}

\end{center}

\vspace{24pt}

The similarity between tree-level string theory scalar amplitudes,
the Koba-Nielsen form ($S^{1}$) and the Virasoro-Shapiro form ($S^{2}$)
suggests a natural $S^{n}$ generalization for a scalar amplitude.
It is shown that the $S^{n}$ amplitude shares many essential properties
of the string theory amplitudes, including $SO(n+1,1)$ conformal symmetry
and linear Regge trajectories for the mass spectrum.
We also discuss factorization and the critical dimension
for the amplitude, which are the necessary conditions for
the quantum mechanical consistency (unitarity) of the
amplitude.

\vfill

\newpage
\pagenumber
\baselineskip=20pt %24->20pt

%%%%%%%%%%%%%%%%%%%%%%
\section{Introduction}
%%%%%%%%%%%%%%%%%%%%%%

In bosonic string theory, the tree-level tachyon amplitudes\footnote{
Strictly, what we mean by `tachyon' is the scalar which is
the ground state of the mass spectrum. The distinction
between tachyon and the ground state scalar is necessary
since the amplitude considered in this paper allows
the higher spin tachyons in its excitation spectrum.}
are given by the Veneziano amplitude for open strings
(and the Koba-Nielsen $M$-point generalization;\, KN hereafter) and
the Virasoro-Shapiro (VS) amplitude for closed strings.
These describe the scattering process by $M$ identical tachyonic scalars.
In integral representation, KN is\footnote{
We adopt a space-time metric with signature $(-,+,+,...,+)$
in accord with standard string theory convention.
The slope parameter $\alpha'$ will be chosen to
be $\alpha'=1$ for convenience.}
\[
  A_{KN} = \int_{S^{1}} \prod_{i=1}^{M} dz_{i}\,
                        \pnorm{i<j}{z_{i}}{z_{j}}
\]
and VS is
\[
  A_{VS} = \int_{S^{2}} \prod_{i=1}^{M} d^{2}z_{i}\,
                        \pnorm{i<j}{z_{i}}{z_{j}}.
\]
Note the remarkable resemblance between KN and VS amplitudes.
The only difference is the domain of integration: the domains are
$S^{1}$ (KN) and $S^{2}$ (VS).
The similarity naturally suggests that one should examine
the following possible formula for an amplitude by extending
the integration domain into $S^{n}$:
\begin{equation}
  A_{M} = \int_{S^{n}} \prod_{i=1}^{M} \msr{i} \pnorm{i<j}{\zi}{\zj}.
\end{equation}
The variables $\zi$ are $n$-dimensional vectors
integrated over the sphere $S^{n}$ and
$|\z\,|^{2}$ should be understood as a norm of the vector $\z$.
As shown in section 3, $A_{M}$ expresses a $M$-point scalar amplitude
like the KN and VS amplitudes.

The main purpose of this paper is to study
the symmetry and the unitarity of the amplitude,
which are the first obvious issues to be investigated.
It turns out that $A_{M}$ has a natural generalization
of the conformal symmetry for the string amplitudes;
the algebra of the symmetry is isomorphic to ${\cal SO}(n+1,1)$
if and only if external scalars satisfy a mass-shell condition
(Recall that ${\cal SL}(2,R) \sim {\cal SO}(2,1)$ and
${\cal SL}(2,C) \sim {\cal SO}(3,1)$.).
The unitarity analysis is not completely conclusive
due to the limitation the analysis has,
but our results do not contradict unitarity;
in particular, $A_{M}$ satisfies the factorization condition,
which is a necessary condition for unitarity.
Our proposal of $A_{M}$ does not have a physical motivation,
but the simplicity of the generalization
and its potential relevance to physics are reasons
enough that the amplitude be taken seriously.

Now, one may enquire
whether the amplitude has any relevance to string theory
or the amplitude suggests some generalization of bosonic string theory.
Let us briefly consider the possible physical interpretation
of the amplitude suggested by this particular representaion.
This formula is defined on $S^{n}$,
not on the world-sheet $S^{2}$ as string theories.
It is then plausible to think $A_{M}$ expresses the dynamics
of relativistic membranes or $p$-branes
and it is our conjecture that
the $S^{n}$ amplitude has some relevance to $p$-brane study.
However, it is not a trivial issue to make sense of
this formula on $p$-branes
since the integrand $\norm{\zi}{\zj}$ may be a
specific characteristic of a 2-dimensional world-sheet.
In string theories,
such a polynomial behavior was the direct consequence
of the facts that
the Green's function in 2-dimensions is given by
\[
 \ln(\mu |z_{i}-z_{j}|),
\]
and that the vertex operator transforms as the wave function
under Poincar\'{e} transformations {\it i.\ e.\ }
\[
 V_{tachyon}=\int d^{2}z e^{ik\cdot X}.
\]
Since Green's functions do not have logarithmic behavior
except in 2-dimensions, one needs
a more complicated vertex operator
in order to interpret the scalar amplitude
as the $p$-brane one.
In view of such difficulty,
we first focus our attention only on unitarity analysis
to decide whether $A_{M}$ is physically sensible or not;
delaying its physical interpretation.

In general, unitarity has two possible consequences:
factorization of $M$-point amplitudes, and
no negative norm states in intermediate processes.
The second consequence
reduces to the problem of obtaining the critical dimension
of the theory.
These issues are discussed in Section 3 and 4.
In order to carry out the critical dimension analysis,
we calculate the 4-point amplitude $A_{4}$ in Section 2.
As a by-product, the symmetry of $A_{M}$ is shown.
Also, we find that in gamma function representation,
$A_{4}$ coincides with a 4-point scalar amplitude
proposed by Virasoro\cite{vi}
(which we call Virasoro amplitude.)
through a dual resonance model study.
\footnote{In dual model context, KN and VS amplitudes are not
the most general scalar amplitudes possible
and the Virasoro amplitude is one example
of a generalization.
Historically, Shapiro noticed
the $M$-point integral representation for the case $n=2$
after Virasoro proposed his formula,
thereby called Virasoro-Shapiro amplitudes.}
In Section 4, it is also shown that
the original Virasoro amplitude is not unitary for non-positive $n$.
This result with the fact that
the parameter $n$ originates in $S^{n}$ imply that
the $S^{n}$ is really a physically important object.

%%%%%%%%%%%%%%%%%%%%%%%%%%%%%%%%%%%%%%%%%
\section{Symmetry and the Virasoro amplitude}
%%%%%%%%%%%%%%%%%%%%%%%%%%%%%%%%%%%%%%%%%

The calculation of $A_{4}$ can be accomplished in a
completely analogous way as the KN and VS formulas.
The first step is to identify the symmetry of $A_{M}$
in order to fix the gauge.

$A_{M}$ is invariant under the following
infinitesimal conformal transformations:
\begin{quote}
   translations: $z'^{\mu}=z^{\mu}+\alpha^{\mu}$

   dilatation: $z'^{\mu}=\beta z^{\mu}$

   $O(n)$ rotations: $z'^{\mu}=\epsilon^{\mu \nu}z_{\nu}$

   special conformal transformations:
       $z'^{\mu}=-2(\gamma \cdot z)z^{\mu}+\gamma^{\mu}|z|^{2}$.
\end{quote}
Its finite form is
\begin{equation}
   \delta z^{\mu}= a^{\mu}+b z^{\mu}+e^{\mu \nu}z_{\nu}
   +\frac{z^{\mu}+c^{\mu}|z|^{2}}{ 1+2c \cdot z+|c|^{2}|z|^{2} }.
\end{equation}
The only nontrivial symmetries are dilatation and
special conformal transformations(SCT).
Like the KN and VS amplitudes,
the invariance under those symmetries
is guaranteed once we impose a mass-shell condition on external scalars.

A SCT is an inversion followed by
a translation and another inversion;
\[
\z
\stackrel{I}{\rightarrow}
   \frac{\z}{|z|^{2}}
\stackrel{T}{\rightarrow}
   \frac{\z+\vc\,|z|^{2}}{|z|^{2}}
\stackrel{I}{\rightarrow}
   \frac{\z+\vc\,|z|^{2}}{ 1+2c \cdot z+|c|^{2}|z|^{2} }
\]
(Some vector symbols are omitted
in order to simplify the expressions.).
Hence, one has to only verify the invariance of $A_{M}$
under the inversion instead of the general SCT.
Under an inversion $\zi \rightarrow \zi/|z_{i}|^{2}$,
\[
  |\zi-\zj|^{2} \rightarrow
  \frac{1}{|z_{i}|^{2}} \frac{1}{|z_{j}|^{2}}|\zi-\zj|^{2},
\]
which yields
\begin{equation}
  \pnorm{i<j}{\zi}{\zj} \rightarrow
  \prod_{i} |\zi|^{-2m_{i}^{2}} \pnorm{i<j}{\zi}{\zj},
\end{equation}
where $m_{i}$ is the mass for the $i$th external particles.
The measure transforms as
\begin{equation}
  \msr{i} \rightarrow |\zi|^{2n} \msr{i}.
\end{equation}
Consequently,
\begin{equation}
  A_{M} \rightarrow
  \int \prod_{i} \msr{i} |\zi|^{-2(m_{i}^{2}+n)}
  \pnorm{i<j}{\zi}{\zj},
\end{equation}
which shows $A_{M}$ is invariant under the inversion
if and only if $m_{i}^{2}=-n$ for all $i$.

The invariance under the dilatation now follows automatically;
under a dilatation $\zi \rightarrow b \zi$,
the measure and the integrand pick up a factor
\begin{equation}
 |b|^{ nM+2 \sum_{i<j} k_{i} \cdot k_{j} }=|b|^{M(n+m^{2})},
\end{equation}
which is just unity by the mass-shell condition.

The algebra of the conformal group
is isomorphic to ${\cal SO}(n+1,1)$.
In differential operator forms, $(n\!+\!1)(n\!+\!2)/2$
generators have the following representations:
\begin{eqnarray*}
  L_{+}^{\mu} & = & \frac{\partial}{\partial z_{\mu}}  \\
  L_{0}       & = & z^{\nu}\frac{\partial}{\partial z^{\nu}}  \\
  L_{0}^{\mu \nu} & = & z^{\mu}\frac{\partial}{\partial z_{\nu}}-
                    z^{\nu}\frac{\partial}{\partial z_{\mu}}  \\
  L_{-}^{\mu} & = & (-2z^{\mu} z^{\nu}+\delta^{\mu \nu}|z|^2)
                      \frac{\partial}{\partial z^{\nu}}.
\end{eqnarray*}
Define $L^{\alpha \beta}=-L^{\beta \alpha}$
(Letters from the beginning of the Greek characters run from 1 to $n+2$.)
such that
\begin{eqnarray*}
   L^{\mu \nu}  & = & L_{0}^{\mu \nu} \\
   L^{\mu \; n+1} & = & \frac{1}{2} (L_{+}^{\mu}-L_{-}^{\mu})  \\
   L^{\mu \; n+2} & = & \frac{1}{2} (L_{+}^{\mu}+L_{-}^{\mu})  \\
   L^{n+1 \; n+2} & = & L_{0}.
\end{eqnarray*}
The generators $L^{\alpha \beta}$ satisfy the commutation relations
\begin{equation}
   [L^{\alpha \beta}, L^{\gamma \delta}]=
      -(\eta^{\alpha \gamma} L^{\beta \delta}
       -\eta^{\beta \gamma} L^{\alpha \delta}
       -\eta^{\alpha \delta} L^{\beta \gamma}
       +\eta^{\beta \delta} L^{\alpha \gamma}),
\end{equation}
where $\eta=$diag$(+,+,...,+,-)$ showing the algebra of the
conformal transformations is isomorphic to ${\cal SO}(n+1,1)$.

Since the translations, dilatation,
and SCT consist of
a non-compact quotient group, the integral $A_{4}$ has to be divided by
the volume factor of the quotient group
by a standard gauge fixing procedure.
Corresponding to the dimensionality of the non-compact
space, $2n+1$ coordinates can be fixed, which we take
$\vz{1}=0, z_{2}^{1}$ so that $|\vz{2}|=1$, and $\vz{3}=\infty$.
Using the residual rotational symmetry of the
amplitude to further fix the gauge
$\vz{2}=\hat{e}_{1}=(1, 0,\cdots, 0)$,
one obtains the Jacobian of the transformation:
\begin{equation}
 \frac{\partial(\vz{1}, z_{2}^{1}, \vz{3})}
      {\partial(\vec{\alpha}, \beta, \vec{\gamma})}=|\vz{3}|^{2n},
\end{equation}
which therefore yields
\begin{eqnarray}
 A_{4} & = & \int \msr{4} | \vz{4} |^{2k_{1}k_{4}}
                      | \hat{e}_{1}-\vz{4} |^{2k_{2}k_{4}}  \\
       & = & \int dw \cdots dy \,
              ( w^{2} \! +\cdots+ \! y^{2})
                                       ^{k_{1}k_{4}}
        \left\{ (1\!-\!w)^{2} \! + \! x^{2} \! +\cdots+ \! y^{2} \right\}
                                       ^{k_{2}k_{4}},
\end{eqnarray}
where $\vz{4}=(w, x,\cdots,y)$.
This integral is evaluated by a standard trick \cite{sc}
and the result is the Virasoro amplitude
for positive integer $n$ \cite{vi}:
\begin{equation}
 A_{4}=\pi^{n/2}
      \frac{\Gamma(-\frac{1}{2}\alpha(s))\Gamma(-\frac{1}{2}\alpha(t))
                                         \Gamma(-\frac{1}{2}\alpha(u))}
           {\Gamma(-\frac{1}{2}\alpha(s)-\frac{1}{2}\alpha(t))
            \Gamma(-\frac{1}{2}\alpha(t)-\frac{1}{2}\alpha(u))
            \Gamma(-\frac{1}{2}\alpha(u)-\frac{1}{2}\alpha(s))}.
\end{equation}
Here $s$, $t$, and $u$ are
the conventional Mandelstam variables.
And $\alpha(s)$, {\it etc.\/} are
the Regge trajectory functions satisfying
\begin{equation}
 \alpha(s)=s+\alpha(0)
\end{equation}
with the intercept
of the Regge trajectory $\alpha(0)=n$.
Even though the original expression $A_{M}$ is defined
only for positive integer $n$,
we can now analytically continue $n$
to be any real numbers once we obtain the above expression.
$A_{4}$ therefore reduces to the Virasoro amplitude.

The amplitude exhibits a pole at $s=2r-n$
in the $s$ channel,
where $r$ is a non-negative integer.
The mass spectrum is, therefore, given by $m^{2}=2r-n$,
which means all particle poles lie
on linear Regge trajectories\cite{vi}
(For the leading trajectory, spin $J=2r$. See equation (19).).
In addition to the linear Regge trajectories,
the Virasoro amplitude shares
the following physical features with KN and VS amplitudes\cite{vi}:
\begin{enumerate}
   \item Crossing symmetry.

   \item Superconvergence sum rules.

   \item Regge behavior at asymptotic energies.
\end{enumerate}

Although the Virasoro formula itself has been known
for over 20 years, the amplitude has been little explored so far.
One problem which has limited the study of the Virasoro amplitude
was the lack of the $M$-point scalar generalization,
since the 4-point scalar amplitude itself
can not be a complete solution
for the most general $S$-matrix elements.
On the other hand, the $M$-point scalar amplitude does contain
the complete solution as we will mention in the next section.
Also, the lack of the integral representaion was another
serious problem since it gives clues
about the symmetry of the underlying theory
and about the elementary quantity of the theory
({\it i.\/e.\/} particle, string, {\it etc.}).
Thus, the $S^{n}$ amplitude may shed light on
the Virasoro amplitude.

%%%%%%%%%%%%%%%%%%%%%%%
\section{Factorization}
%%%%%%%%%%%%%%%%%%%%%%%

We now examine the unitarity of the amplitude.
Since our knowledge is limited
only to amplitude formulas and
the underlying theory is unknown,
we lack the systematic analyses
using operator formalism or path integral approach,
which proved their powers in string theory
to show factorization and critical dimension.
In order to carry out these analyses,
we employ the methods which do not rely on these formalisms
but rely only on the amplitude formula.

We employ the same method discussed by Mandelstam\cite{ma}
to show factorization.

In tree-level unitarity,
factorization requires the following.
Consider a tree-level process with $M$ scalars.
\begin{figure}[t]
  \vspace{2.5in}
  \caption{Factorization requires that the residue of the
           $1 2 \cdots m \rightarrow (m+1) \cdots M$ channel
           should be the product of $1 2 \cdots m P$ and
           $P m+1 \cdots M$ tree amplitudes.}
\end{figure}
The residue of a pole associated with
$1 2 \cdots m \rightarrow (m+1) \cdots M$ channel
should be the product of three factors.
The first and the second factors are
tree-level subprocess amplitudes with $1 2 \cdots m P$
and $P (m+1) \cdots M$ respectively.
The last factor is an angular factor
depending on the angular momentum of the intermediate state
(Equation (19) gives a simple example of factorization.).
For a scalar pole, the angular factor is a numerical factor,
therefore the residue must be the product of
two tree-amplitude formulas,
which are $m+1$ and $M-m+1$ scalar amplitudes.

The analysis is of course important by itself for unitarity,
but there is another reason why we would like to
stress this study.
The factorization of $A_{M}$
implies only tree-level unitarity
if $A_{M}$ is regarded as a Born term
in a perturbation expansion
as KN and VS amplitudes\cite{ki1}.
But once one proves this,
factorization enables one to formulate the loop amplitudes
required for full unitarity.
Moreover, the general $S$-matrix elements are now constructed
by the repeated factorization
from the $M$-point scalar amplitude\cite{ma}.

The fixed variables for the gauge fixing are chosen to be
$\vz{1}=0$, \linebreak
$\vz{m+1}=\hat{e}_{1}$, and $\vz{M}=\infty$.
We also introduce the polar coordinate for $\vz{m}$,
$\vz{m}=(\rho, \phi, \theta_{1},\cdots, \theta_{n-2})$
in which $0 < \rho < \infty, 0< \phi < 2 \pi,$
and $0 < \theta_{k} < \pi$.

Now, define new variables by
\begin{equation}
  \pzi=\frac{1}{\rho} \Rt \zi
\end{equation}
for particles $i=1,2, \cdots ,m$.
Here, the rotation matrix
$\Rt(\phi, \theta_{k})$ is defined so that
$\Rt \vz{m}=\rho \hat{e}_{1}$.
Note $\vec{z}\,'_{m}=\hat{e}_{1}$.

In terms of the new variables, $A_{M}$ is expressed as
\begin{eqnarray}
  A_{M} & = & \int_{0}^{\infty} d\rho\, \rho^{-s+m^{2}+1}
              \int d\phi
              \prod_{k=1}^{n-2} \sin^{k}\theta_{k} d\theta_{k}
                                                    \nonumber \\
        &   & \int \pmsr{2} \cdots \pmsr{m-1}
                   \msr{m+2} \cdots \msr{M-1} \nonumber \\
        &   & \pnorm{i<j \leq m}{\pzi}{\pzj}
              \pnorm{i>j \geq m}{\zi}{\zj}
              \pnorm{\stackrel{i>m}{j \leq m}}
                                {\zi}{\rho R \pzj},\mbox{}
\end{eqnarray}
where $s=-(k_{1}+ \cdots +k_{m})^{2}$.

We expand the last factor in a Taylor series around $\rho=0$:
\begin{equation}
  \pnorm{i>m, j \leq m}{\zi}{\rho R \pzj} =
        \sum_{r} \frac{1}{r!}
        \left.
         \left( \frac{\partial}{\partial \rho} \right)^{r}
         \pnorm{i>m, j \leq m}{\zi}{\rho R \pzj}
        \right|_{\rho=0} \rho^{r}.
\end{equation}
The $\rho$ integration in (14) gives a pole
at $s=2r-n$ for $n>1$ or at $s=r-n$ for $n=1$.
This follows by observing that
\begin{equation}
  |\zi-\rho R \pzj|^{2} =
        |\zi|^{2}-2\rho \zi \cdot (R \pzj)+\rho^{2}|\pzj|^{2}.
\end{equation}
Consider the terms in (15) which contain the odd power of $r$;
those are proportional to
the second term in (16)
which vanishes by the angular part of the integral in (14).
On the other hand, the even $r$ terms always contain
the terms which are angle independent, hence
nonvanishing by the integral. Obviously,
this is the case except $n=1$.

The residue $R_{2r}$ for the pole at $s=2r-n$ is
\begin{eqnarray}
  R_{2r} & \propto &  \int \pmsr{2} \cdots \pmsr{m-1}
                 \pnorm{i<j \leq m}{\pzi}{\pzj} \nonumber \\
         &         &  \mbox{} \times \int \msr{m+2}
                                          \cdots \msr{M-1}
                 \pnorm{i>j \geq m}{\zi}{\zj}
                 \prod_{i>m} | \zi |
                           ^{-2k_{i} \sum_{j>m} k_{j}} \nonumber \\
         &         &  \mbox{} \times F_{2r}(\zi, \pzj, k_{i}k_{j}),
\end{eqnarray}
where $F_{2r}$ is the angular factor.
This factor can be decomposed into a sum of terms,
each consisting of two factors
which depend only on $\z$ and $\pz$ respectively;
therefore implies factorization.
Each term in the sum expresses the resonance
with spin ranging from 0 to $2r$.

As an explicit example,
the residue $R_{0}$ associated with the first pole, $s=-n$, is
\begin{eqnarray}
  R_{0} & \propto &  \int \pmsr{2} \cdots \pmsr{m-1}
                 \pnorm{i<j \leq m}{\pzi}{\pzj} \nonumber \\
   &          &  \times \int \msr{m+2} \cdots \msr{M-1}
                 \pnorm{i>j \geq m}{\zi}{\zj}
                 \prod_{i>m} | \zi |^{-2k_{i} \sum_{j>m} k_{j}}.
\end{eqnarray}
The residue is separated into two integrals,
the integrals with $i<j \leq m$ variables and
the integrals with $i>j \geq m$ variables.
Thus, we can identify these as $A_{m+1}$ and $A_{M-m+1}$
respectively with the sets of fixed variables:
\[
\left \{ \begin{array}{llll}
 A_{m+1}:   & \pvz{1}=0,& \pvz{m}=\hat{e}_{1}, & \pvz{P}=\infty \\
 A_{M-m+1}: & \vz{P}=0, & \vz{m+1}=\hat{e}_{1},& \vz{M}=\infty.
         \end{array} \right.
\]
Here, $\vz{P}$ and $\pvz{P}$ are the new variables
corresponding to the particle in the intermediate states.

This example gives a proof that the external tachyons are scalars,
and the same scalars as the ground state
on the leading Regge trajectory.

%%%%%%%%%%%%%%%%%%%%%%%%%%%%
\section{Critical Dimension}
%%%%%%%%%%%%%%%%%%%%%%%%%%%%

Since $A_{M}$ describes a spinless scattering process
as shown above,
factorization and partial wave analysis demand that
the residue $R$ of the 4-point amplitude $A_{4}$ is expressed as
\begin{equation}
  R = \sum_{l=0}^{\infty} G_{l}^{2} P_{l}(z)
\end{equation}
for the incoming and outgoing states
which differ by a relative angle. Here, $z$ is
$\cos \theta$, where $\theta$ is
center-of-mass scattering angle.
$P_{l}$ are Legendre polynomials
in $d$-dimensional spacetime\cite{ho}.
Up to numerical factors,
$G_{l}$'s are the coupling constants of the external scalars
with intermediate spin-$l$ particles.
In general, the hermitity of a Lagrangian,
which requires the couplings to be real,
also implies $G_{l}^{2} \geq 0$.
This requirement strongly constrains
a given amplitude formula, so that the formula is valid
only for a small interval of $d$.

This is a nice trick to get the magic number 26 for open string\cite{fr}.
Consider the second pole
in the Veneziano amplitude, namely the $\alpha(s)=2$ pole.
The residue is given by
\begin{equation}
  R_{2}(z) \propto (z^{2}-\frac{1}{25}).
\end{equation}
The corresponding partial wave analysis formula gives
\begin{equation}
  R_{2}(z)= G_{2}^{2} (z^{2}-\frac{1}{d-1})+G_{0}^{2}.
\end{equation}
Comparing these two formulas, one concludes
the coupling of the scalar, $G_{0}^{2}$,
is negative for $d>26$.
The scalar completely decouples and the intermediate state
becomes pure spin-2 when $d=26$.

The problem of the method is that this does not work
even for closed strings.
The first massive level for closed string is
$\alpha(s)=4$ pole with $s=2$.
A similar calculation gives $d=72$,
which is certainly wrong.

The origin of the problem is not difficult to see.
A closed string state is formed by a tensor product of an
open string state with itself.
The open string state for $\alpha(s)=2$
contained a physical spin-2 and an unphysical spin-0 state.
So, the corresponding closed string state $\alpha(s)=4$
contains a physical spin-2 state
in addition to two unphysical spin-2 states.
Thus, the amplitude with the incoming and outgoing states
$|s\rangle$ and $|s'\rangle$, is written as
\[
   \langle s'|s \rangle= \cdots +
%|\langle s|4 \rangle|^{2} P_{4}(z) +
                 \left\{ (26-d)| \langle s|u_{1}\rangle |^{2} +
                         (26-d)| \langle s|u_{2}\rangle |^{2} +
                               | \langle s|p\rangle |^{2}
                   \right\} P_{2}(z) +
                                 \cdots,
\]
where $|u_{i} \rangle$ and $|p\rangle$ are unphysical and
physical spin-2 states respectively.
Also, we extract the angular dependence from the amplitude.
Since $|\langle s|p\rangle|^{2}$ is positive definite,
the value within the braces can be positive even when
$d>26$. In other words, the problem is that
there exist several distinct states at a given spin level,
so that the amplitude must be described
not by a single coupling $G_{2}^{2}$,
but by several $(G_{2}^{i})^{2}$, which have to be
{\it all} positive in order to satisfy unitarity.

Supposing such physical states
also `contaminate' $A_{4}$ for general $n$, one gets
the necessary condition for unitarity,
but not the sufficient condition.
For a $\alpha(s)=2r$ pole,
where $r$ is a positive integer, we obtain
\begin{equation}
   d \leq 5 - 4r + \frac{r(2r-1)(3n+2r)^{2}}
                  {\sum_{i=0}^{r-1} (-n+2r-4i)^{2}}
\end{equation}
by demanding $(G_{2r-2})^{2} \geq 0$.
The lower spin state equations, $(G_{2r-2i})^{2} \geq 0$,
where $1 < i \leq r$, gives the weaker conditions
on $d$ in general.
For closed string case, this is because the lower spin levels
contain more and more physical states.
Eq.\ (22) gives
$d<57$ and 83 for $S^{2}$ and $S^{3}$ respectively, at their minima.
In spite of the limitation the analysis has, eq.\ (22)
indicates that the Virasoro amplitude for non-positive
$n$ does not lead to sensible quantum theories.
$n=0$ gives $d<2$
and negative $n$ have negative critical dimensions.
This exclusion of the non-positive region for $n$
might imply $S^{n}$ is not simply
a convenient integral representation
but really a physical object
such that non-positive $n$ are ill-defined.

%%%%%%%%%%%%%%%%%%%%
\section{Comments}
%%%%%%%%%%%%%%%%%%%%

In this paper, we explored some aspects of the $S^{n}$ amplitude
putting stress on unitarity.
In the future, the question of critical dimension
must be regarded as
the first and foremost problem to be solved.

There is one possibility which may improve our calculation
of critical dimension. Consider more general configurations
than 2-body scattering, such as 3-body scattering.
The residues can be evaluated in the same manner as in
Section 3, but now the residue of the poles contain
the particle four-momenta as free parameters which we can vary.
Then, one might find the region in parameter space in which
the physical states decouple.

Another problem we must solve is to derive $A_{M}$ as a $p$-brane
amplitude. We have noticed this is a non-trivial issue, but
it becomes more evident if one notes that $A_{M}$ has
the linear Regge trajectories.
This clearly contradicts the relation
given by Kikkawa and Yamasaki\cite{ki2}.
The difference comes from the fact that (1)
needs a dimensionful constant $\alpha'$ in the exponent;
therefore, the `tension' has the unit $M/L$ whereas
the $p$-brane tension has the unit $M/L^{p}$.

Unfortunately, this amplitude is `worse' than the conventional
string theories: the mass spectrum is more and more `tachyonic'
as $n$ increases since the Virasoro amplitude gives
$m^{2}=J-n$. On the other hand, this means one gets
massless states with higher spins.
For example, the $S^{4}$ case contains
a spin-4 massless state.
This is a point worth making
since no quantum mechanically consistent theories
with massless high spin particles are known.
Our amplitude therefore may provide a theory
with massless high spin particles
just as string theories provided consistent theories with
massless spin-2 particles for the first time in physics history.

\vspace{.1in}

\begin{sloppypar}

\begin{center}
    {\Large {\bf Note added} }
\end{center}
\vspace{.1in}

After the completion of this paper, I learned there exist papers
which cover some of this work.
For the earliest work, see R. C. Brower and P. Goddard, Lett. Nuovo Cimento
{\bf 1}, 1075 (1971).
A recent work on this idea is M. B. Green and C. B. Thorn, Nucl. Phys.
{\bf B367}, 462 (1991).
I wish to thank D. Fairlie and M. B. Green for comments about these
early literatures.

\begin{center}
    {\Large {\bf Acknowledgements} }
\end{center}
\vspace{.1in}

I am grateful to J. Polchinski for having suggested the problem
and for his continuous assistance throughout the work. I also
thank J. LaChapelle for critical reading of the manuscript.
This research was supported in part by the Robert A. Welch
Foundation, NSF Grant PHY 8904035 and 9009850, and
the Texas Advanced Research Program Grant 476.

\end{sloppypar}

\pagebreak

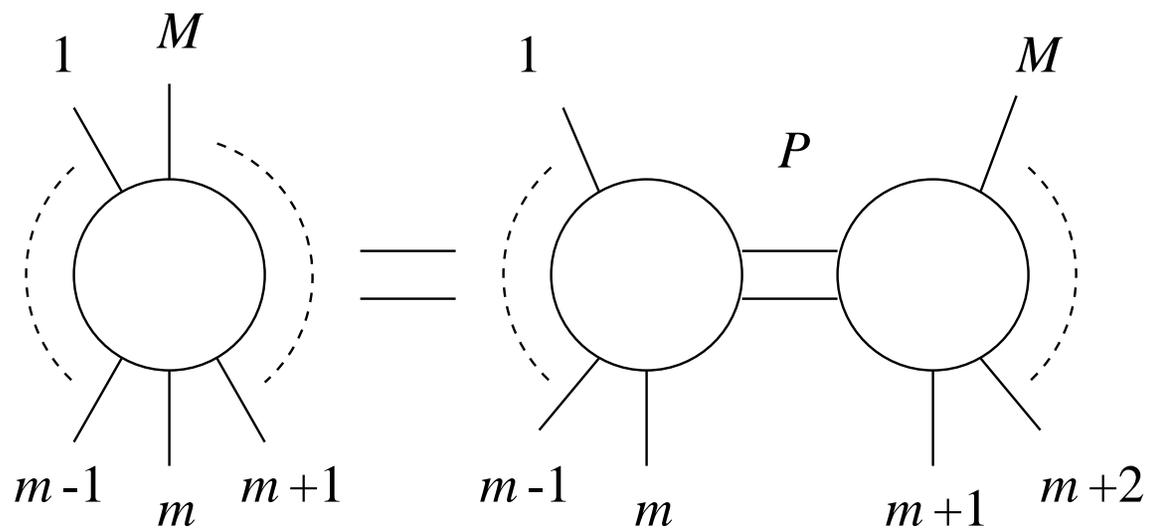
\end{document}